\documentclass[conference]{IEEEtran}
\usepackage[boxruled]{algorithm2e}
\usepackage{cite}
\usepackage{graphicx}
\usepackage{psfrag}
\usepackage{subfigure}
\usepackage{url}
\usepackage{amsmath}
\usepackage{array}
\usepackage{amssymb}
\usepackage{amsfonts}
\usepackage{graphicx}
\usepackage{epstopdf}
\usepackage{algorithmic}

\newtheorem{lemma}{Lemma}

\newtheorem{definition}{Definition}
\newtheorem{theorem}{Theorem}

\newtheorem{example}{Example}

\usepackage{setspace}
\usepackage{amscd}
\usepackage{mathrsfs}
\usepackage{epsfig}
\usepackage{color}
\usepackage{textcomp}
\usepackage{multirow}
\usepackage{threeparttable}

\title{On the Vector Linear Solvability of Networks and Discrete Polymatroids}
\begin{document}

\author{
\authorblockN{Vijayvaradharaj T. Muralidharan and B. Sundar Rajan}
\authorblockA{Dept. of ECE, Indian Institute of Science, Bangalore 560012, India, Email:{$\lbrace$tmvijay, bsrajan$\rbrace$}@ece.iisc.ernet.in
}
}

\maketitle
\begin{abstract}
We consider the vector linear solvability of networks over a field $\mathbb{F}_q.$ It is well known that a scalar linear solution over $\mathbb{F}_q$ exists for a network  if and only if the network is \textit{matroidal} with respect to a \textit{matroid} representable over $\mathbb{F}_q.$  A \textit{discrete polymatroid} is the multi-set analogue of a matroid. In this paper, a \textit{discrete polymatroidal} network is defined and it is shown that a vector linear solution over a field $\mathbb{F}_q$ exists for a network if and only if the network is discrete polymatroidal with respect to a discrete polymatroid representable over $\mathbb{F}_q.$ An algorithm to construct networks starting from a discrete polymatroid is provided. Every representation over $\mathbb{F}_q$ for the discrete polymatroid, results in a vector linear solution over $\mathbb{F}_q$ for the constructed network.  Examples which illustrate the construction algorithm are provided, in which the resulting networks admit vector linear solution but no scalar linear solution over $\mathbb{F}_q.$ 
\end{abstract}
\section{Introduction and Background}
The concept of network coding, originally introduced by Ahlswede et. al. in \cite{Ah}, helps towards providing more throughput in a communication network than what pure routing solutions provide. For multicast networks, it was shown in \cite{Li} that linear solutions exist for sufficiently large field size. An algebraic framework for finding linear solutions in networks was introduced in \cite{KoMe}.

The connection between matroids and  network coding was first established by Dougherty et. al. in \cite{DoFrZe}. In \cite{DoFrZe}, the notion of \textit{matroidal network} was introduced and it was shown that if a scalar linear solution over $\mathbb{F}_q$ exists for a network, then the network is matroidal with respect to a matroid representable over $\mathbb{F}_q.$ The converse that a scalar linear solution exists for a network if the network is matroidal with respect to a matroid representable over $\mathbb{F}_q$ was shown in \cite{KiMe}. 

A construction procedure was given in \cite{DoFrZe} to obtain networks from matroids, in which the resulting network admits a scalar linear solution over $\mathbb{F}_q,$ if the matroid is representable over $\mathbb{F}_q.$ Using the networks constructed using the construction procedure given in \cite{DoFrZe}, it was shown in \cite{DoFrZe_In} that there exists networks which do not admit any scalar and vector linear solution over $\mathbb{F}_q,$ but admit non-linear solution over $\mathbb{F}_q.$ 

In \cite{RoSpGe}, it was shown that an instance of the network coding problem $\mathcal{N}$ can be reduced to an instance of the index coding problem $\mathcal{I}_\mathcal{N}$ and a vector linear solution exists for $\mathcal{N}$ if and only if a particular class of index coding solutions called the \textit{perfect linear index coding} solution exists for $\mathcal{I}_\mathcal{N}.$ Also, in \cite{RoSpGe}, in terms of the circuits and basis sets of a matroid $\mathbb{M},$ an instance of the index coding problem $\mathcal{I}_{\mathbb{M}}$ was defined and it was shown that a perfect linear index coding solution exists for $\mathcal{I}_{\mathbb{M}}$ if and only if $\mathbb{M}$ has a multi-linear representation. 

Extending the notion of matroidal network to networks which admit error correction, it was shown in \cite{PrRa_ISIT} that a network admits a scalar linear
error correcting network code if and only if it is a matroidal error correcting network associated with a representable matroid. Constructions of networks from matroids with correction capability were provided in \cite{PrRa_ISIT,PrRa_ISITA}. 

Discrete polymatroids, introduced by  Herzog and Hibi in \cite{HeHi}, are the multi-set analogue of matroids.  In this paper, we establish the connection between vector linear solvability of networks and the representation of discrete polymatroids. The contributions of this paper are as follows:
\begin{itemize}
\item
The notion of \textit{discrete polymatroidal network} is introduced, which is a generalization of the notion of matroidal network introduced in \cite{DoFrZe}. It is shown that a vector linear solution exists for a network over a field $\mathbb{F}_q$ if and only if the network is discrete polymatroidal with respect to a discrete polymatroid representable over $\mathbb{F}_q.$
\item
An algorithm to obtain networks from a discrete polymatroid is provided. Starting from a discrete polymatroid which is representable over $\mathbb{F}_q,$ the resulting networks admit a vector linear solution over $\mathbb{F}_q.$
\item
Sample constructions of networks obtained from discrete polymatroids which admit a vector linear solution over $\mathbb{F}_q$ but no scalar linear solution over $\mathbb{F}_q$ are provided.    
\end{itemize}
\textbf{\textit{Notations:}}
The set $\lbrace 1,2,\dotso,n \rbrace$ is denoted as  $\lceil n \rfloor.$ $\mathbb{Z}_{\geq 0} $ denotes the set of non-negative integers. For a vector $v$ of length $n$ and $A \subseteq \lceil n \rfloor,$ $v(A)$ is the vector obtained by taking only the components of $v$ indexed by the elements of $A.$ The vector whose $i^{\text{th}}$ component is one and all other components are zeros is denoted as $\epsilon_{i}.$  For $u,v \in \mathbb{Z}_{\geq 0}^n,$ $u \leq v$ if all the components of $v-u$ are non-negative and,  $u < v$ if $u \leq v$ and $u \neq v.$ For $u,v \in \mathbb{Z}_{\geq 0}^n,$ $u \vee v$ is the vector whose $i^{\text{th}}$ component is the maximum of the $i^{\text{th}}$ components of $u$ and $v.$ A vector $u \in \mathbb{Z}_{\geq 0}^n$ is called an integral sub-vector of $v \in \mathbb{Z}_{\geq 0}^n$ if $u < v.$ For a set $A,$ $\vert A \vert$ denotes its cardinality and for a vector $v \in  \mathbb{Z}_{\geq 0}^n,$ $\vert v \vert$ denotes the sum of the components of $v.$ 
\section{Preliminaries}
\subsection{Network Coding}
In this subsection, the basic definitions and notations related to networks and their solvability are defined.  

A communication network is a directed, acyclic graph with the set of vertices denoted by $\mathcal{V}$ and the set of edges denoted by $\mathcal{E},$ with $\vert \mathcal{E}\vert = l.$ All the edges in the network are assumed to have unit capacity over $\mathbb{F}_q^k,$ i.e., they can carry a vector of dimension $k$ over $\mathbb{F}_q.$ The in-degree of an edge $e$ is the in-degree of its head vertex and out-degree of $e$ is the out-degree of its tail vertex. The messages in the network are generated at edges with in-degree zero, which are called the input edges of the network and let $\mathcal{S}  \subset \mathcal{E},$ denote the set of input edges with $\vert \mathcal{S} \vert =m.$ The edges other than the input edges are referred to as the intermediate edges. A vertex $v \in \mathcal{V}$ demands the set of messages generated at the input edges given by $\delta(v) \subseteq \mathcal{S},$ where $\delta$ is called the demand function of the network. $In(v)$ denotes the set of incoming edges of a vertex $v $ ($In(v)$ includes the intermediate edges as well as the input edges which are incoming edges at node $v$) and $Out(v)$ denotes the union of the set of intermediate edges emanating from $v$ and $\delta(v).$
For an intermediate edge $e,$ $head(e)$ and $tail(e)$ respectively denote the head vertex and tail vertex of $e.$ 
  
An edge carries a vector of dimension $k$ over $\mathbb{F}_q.$ Let $x_i, i \in \lceil m \rfloor,$ denote the vector generated at the $m$ input edges of the network. Let $x=[ x_1,x_2,\dotso,x_m ].$

A vector network code of dimension $k$ over $\mathbb{F}_q$ is a collection of functions $\lbrace \psi_e : \mathbb{F}_q^{km} \rightarrow {\mathbb{F}_q}^{k} , e \in \mathcal{E}\rbrace,$ where the function $\psi_e$ is called the global encoding function associated with the edge $e.$  The global encoding functions satisfy the following conditions:
\begin{description}
\item [(N1):]
$\psi_i(x)=x_i, \forall i \in \mathcal{S},$
\item [(N2):]
For every $v \in \mathcal{V},$ for all $j \in \delta(v),$ there exists a function $\chi_{v,j} : \mathbb{F}_{q}^{k\vert In(v)\vert} \rightarrow \mathbb{F}_q^k$ called the decoding function for message $j$ at node $v$ which satisfies $\chi_{v,j}(\psi_{i_1}(x),\psi_{i_2}(x), \dotso, \psi_{i_t(x)})=x_{j},$  where $In(v)=\{i_1,i_2, \dotso i_t \} .$
\item [(N3):]
For all \mbox{$i \in \mathcal{E} \setminus \mathcal{S},$} there exists \mbox{$\phi_i: {\mathbb{F}_q}^{k \vert In(head(i))\vert} \rightarrow \mathbb{F}_q^k$} such that \mbox{$\psi_i(x)=\phi_i(\psi_{i_1}(x),\psi_{i_2}(x), \dotso, \psi_{i_r}(x)),$}  where $In(head(i))=\{i_1,i_2, \dotso i_r \}.$ The function $\phi_i$ is called the local encoding function associated with edge $i.$ 
\end{description}

A network coding solution with $k=1$ is called a scalar solution; otherwise the solution is a vector solution. A solution for which all the local encoding functions and hence the global encoding functions are linear is said to be a linear solution. A network for which a solution (scalar linear solution/ vector linear solution) exists is said to be solvable (scalar linear solvable/ vector linear solvable). For a vector linear solution, the global encoding function $\psi_i, i \in \mathcal{E},$ is of the form $\psi_i(x)= x M_i,$ where $M_i$ is an $mk \times k$ matrix over $\mathbb{F}_q$ called the global encoding matrix associated with edge $i.$
\subsection{Discrete Polymatroids and Matroids}
In this subsection, a brief overview of the concepts related to discrete polymatroids and matroids and their representability is presented. For a comprehensive treatment, interested readers are referred to \cite{We,Ox,HeHi,Vl}. The notion of a discrete polymatroidal network is introduced in the next section, which is a generalization of the notion of a matroidal network introduced in \cite{DoFrZe}.
\subsubsection{Discrete Polymatroids}
\begin{definition}[\cite{HeHi}]
Let $\mathbb{D}$ be a non-empty finite set of vectors in $\mathbb{Z}_{\geq 0}^n,$
which contains with each $u \in \mathbb{D}$ all its integral sub-vectors. The set $\mathbb{D}$ is called
a discrete polymatroid on the ground set $\lceil n \rfloor$ if for all $u, v \in \mathbb{D}$ with $\vert u\vert < \vert v\vert,$
there is a vector $w \in  \mathbb{D}$ such that $u < w  \leq  u \vee v.$
\end{definition}

The function $\rho^{\mathbb{D}}: 2^{\lceil n \rfloor} \rightarrow \mathbb{Z}_{\geq 0}$ called the rank function of $\mathbb{D}$ is defined as $\rho^{\mathbb{D}}(A)=\max \{ \vert u(A) \vert , u \in \mathbb{D}\},$ where $\phi \neq A  \subseteq \lceil n \rfloor$ and $\rho^{\mathbb{D}}(\phi)=0.$ In terms of the rank function $\rho^{\mathbb{D}},$ the discrete polymatroid can be written as  $\mathbb{D}=\lbrace x \in \mathbb{Z}_{\geq 0}^n: \vert x(A) \vert \leq \rho^{\mathbb{D}}(A), \forall A \subseteq \lceil n \rfloor \rbrace.$  For simplicity, in the rest of the paper, the rank function of $\mathbb{D}$ is denoted as $\rho.$

From Proposition 4 in \cite{FaFaPa}, it follows that the a function $\rho: 2^{\lceil n \rfloor} \rightarrow \mathbb{Z}_{\geq 0}$ is the rank function of a discrete polymatroid if and only if it satisfies the conditions,
\begin{description}
\item [(D1)]
If $A \subseteq B \subseteq \lceil n \rfloor,$ then $\rho(A)\leq \rho(B).$
\item [(D2)]
 $\forall A,B \subseteq \lceil n \rfloor,$  $\rho(A \cup B) + \rho(A\cap B)\leq\rho(A)+\rho(B).$
\item [(D3)]
$\rho(\phi)=0.$
\end{description}
 
 A vector $u \in \mathbb{D}$ is a basis vector of $\mathbb{D},$ if $u <v$ for no $v \neq u \in \mathbb{D}.$ The set of basis vectors of $\mathbb{D}$ is denoted as $\mathcal{B}(\mathbb{D}).$ For all $u \in \mathcal{B}(\mathbb{D}),$ $\vert u \vert$ is equal \cite{Vl}, which is called the rank of $\mathbb{D},$ denoted by $rank(\mathbb{D}).$  
\begin{example}
\begin{figure}[htbp]
\centering
\includegraphics[totalheight=2.5 in,width=3.5in]{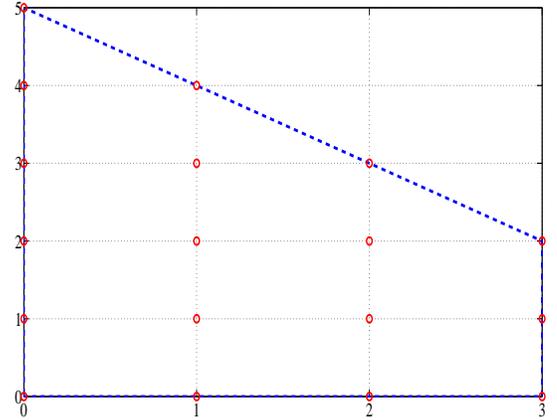}
\caption{An example of a discrete polymatroid}
\label{example_2d_ind}
\end{figure}
\label{example_2d}
Let $\rho: 2^{\lceil 2 \rfloor} \rightarrow \mathbb{Z}_{\geq 0}$ be defined as follows:
$\rho (\{1\})=3$ and $\rho(\{2\})=\rho(\{1,2\})=5.$ It can be seen that $\rho$ satisfies (D1)--(D3) and hence  $\rho$ is the rank function of a discrete polymatroid. The vectors which belong to this discrete polymatroid are the points marked by red in Fig. \ref{example_2d_ind}. The set of basis vectors for this discrete polymatroid is given by $\{(0,5),(1,4),(2,3),(3,2)\}.$
\end{example}
\begin{example}[\cite{Vl}]
Let \mbox{$\rho: 2^{\lceil 3 \rfloor} \rightarrow \mathbb{Z}_{\geq 0}$} be defined as follows: $\rho(\phi)=0,$ $\rho(\{1\})=1,$ $\rho(\{2\})=\rho(\{3\})=\rho(\{1,3\})=2,$ $\rho(\{1,2 \})=3,$ $\rho(\{2,3\})=\rho(\{1,2,3\})=4.$ It can be verified that $\rho$ satisfies (D1)--(D3) and hence is the rank function of the discrete polymatroid given by,

{\footnotesize
\begin{align*}
\{&(0, 0, 0), (1, 0, 0), (0, 1, 0), (0, 0, 1), (1, 1, 0), (0, 1, 1), (1, 0, 1), (0, 2, 0),\\
 &\hspace{1 cm}(0, 0, 2),(0, 1, 2), (0, 2, 1), (1, 1, 1), (1, 2, 0), (0, 2, 2), (1, 2, 1)\}.
\end{align*}
} 
\noindent The set of basis vectors of this discrete polymatroid is $\mathcal{B}(\mathbb{D})=\{(0,2,2),(1,2,1)\}.$
\endproof
\end{example}

\begin{example}
\label{example1}
Consider the function $\rho : 2^{\lceil 4 \rfloor} \rightarrow \mathbb{Z}_{\geq 0}$ defined as
\begin{displaymath}
   \rho(X)= \left\{
     \begin{array}{lr}
       2\vert X \vert & :\text{if\;} \vert X \vert \leq 2\\
       4 & : \text{otherwise\;}
     \end{array}.
   \right.
\end{displaymath} 
It can be verified that $\rho$ satisfies (D1)--(D3). The set of basis vectors of the discrete polymatroid $\mathbb{D}$ of which $\rho$ is the rank function is given by,

{\footnotesize
\begin{align*}
&\left \lbrace (0,0,2,2),(0,1,1,2),(0,1,2,1),(0,2,0,2),(0,2,1,1),(0,2,2,0),\right.\\
&\hspace{0.2 cm}(1,0,1,2),(1,0,2,1),(1,1,0,2),(1,1,1,1),(1,1,2,0),(1,2,0,1),\\
&\hspace{0.2 cm}(1,2,1,0),(2,0,0,2),(2,0,1,1),(2,0,2,0),(2,1,0,1),(2,1,1,0),\\&\hspace{6.9 cm}\left.(2,2,0,0)\right\}.
\end{align*}
}
\endproof
\end{example}

Let $E$  be a vector space over $\mathbb{F}_q$ and $V_1,V_2,\dotso, V_n$ be finite dimensional vector subspaces of $E.$ Let the mapping $\rho: 2^{\lceil n \rfloor} \rightarrow \mathbb{Z}_{\geq 0}$ be defined as $\rho(X)=dim(\sum_{i \in X} V_i), X \subseteq \lceil n \rfloor.$ It can be verified that $\rho$ satisfies (D1)--(D3) and is the rank function of a discrete polymatroid, denoted by $\mathbb{D}(V_1,V_2,\dotso,V_n).$ 
Note that $\rho$ remains the same even if we replace the vector space $E$ by the sum of the vector subspaces $V_1,V_2,\dotso,V_n.$ In the rest of the paper, the vector subspace $E$ is taken to be the sum of the  vector subspaces $V_1,V_2,\dotso,V_n$ considered. 
The vector subspaces $V_1,V_2,\dotso,V_n$ can be described by a matrix $A=[A_1 \; A_2 \; \dotso A_n ],$ where $A_i, i \in \lceil n \rfloor,$ is a matrix whose columns span $V_i.$

\begin{definition}[\cite{FaFaPa}]
A discrete polymatroid $\mathbb{D}$ is said to be representable over $\mathbb{F}_q$ if there exists vector subspaces $V_1,V_2,\dotso,V_n$ of a vector space $E$ over $\mathbb{F}_q$ such that $dim(\sum_{i \in X} V_i)=\rho(X), \forall X \subseteq \lceil n \rfloor.$ The set of vector subspaces $V_i,i\in\lceil n \rfloor,$ is said to form a representation of $\mathbb{D}.$
\end{definition}

\begin{example}
Consider the discrete polymatroid $\mathbb{D}$ given in Example \ref{example1}.
Let $A= \underbrace{\hspace{-.2 cm}\left[\begin{matrix} 1&0\\ 0&1\\0&0\\0&0 \end{matrix}\right.}_{A_1} \quad \underbrace{\begin{matrix} 0&0\\ 0&0\\1&0\\0&1 \end{matrix}}_{A_2} \quad \underbrace{\begin{matrix} 1&0\\ 0&1\\0&1\\1&0 \end{matrix}}_{A_3}  \quad \underbrace{\left.\begin{matrix} 1&0\\ 0&1\\1&0\\1&1 \end{matrix}\right]}_{A_4}$ be a matrix over $\mathbb{F}_2.$
Let $V_i$ denote the column span of $A_i, i \in \lceil 4 \rfloor.$ It can be verified that the vector subspaces $V_1,$ $V_2$ and $V_3$ and $V_4$ form a representation of the discrete polymatroid $\mathbb{D}$ over $\mathbb{F}_2.$
\endproof
\end{example}
\subsubsection{Matroids}
\begin{definition}[\cite{We}]
A matroid is a pair $(\lceil n \rfloor, \mathcal{I}),$ where $\mathcal{I}$ is a collection of  subsets of $\lceil n \rfloor$ satisfying the following three axioms:
\begin{itemize}
\item
$\phi \in \mathcal{I}.$
\item
If $X \in \mathcal{I}$ and $Y \subseteq I,$ then $Y \in \mathcal{I}.$
\item
If $U,V$ are members of $\mathcal{I}$ with $\vert U \vert =\vert V \vert +1$ there exists $x \in U \setminus V$ such that $V \cup x \in \mathcal{I}.$
\end{itemize} 
\end{definition}

A subset of $\lceil n \rfloor$ not belonging to $\mathcal{I}$ is called a dependent set. A maximal independent set is called a basis set and a minimal dependent set is called a circuit. The rank function of a matroid $r: 2^{\lceil n \rfloor} \rightarrow \mathbb{Z}_{\geq 0}$ is defined by
$r(A)=\max\{\vert X \vert: X \subseteq A, X \in \mathcal{I} \},$ where $A \subseteq \lceil n \rfloor.$ The rank of the matroid $\mathbb{M},$ denoted by $rank(\mathbb{M})$ is equal to $r(\lceil n \rfloor).$

 A function $r: 2^{\lceil n \rfloor} \rightarrow \mathbb{Z}_{\geq 0}$ is the rank function of a matroid if and only if it satisfies the conditions (D1)--(D3) and the additional condition that $r(X)\leq \vert X \vert, \forall X \subseteq \lceil n \rfloor$ (follows from Theorem 3 in Chapter 1.2 in \cite{We}). Since the rank function $r$ of $\mathbb{M}$ satisfies (D1)--(D3), it is also the rank function of a discrete polymatroid denoted as $\mathbb{D}(\mathbb{M}).$ In terms of the set of independent vectors $\mathcal{I}$ of $\mathbb{M},$  the discrete polymatroid $\mathbb{D}(\mathbb{M})$ can be written as $\mathbb{D}(\mathbb{M})=\{\sum_{i \in I} \epsilon_{i,n} :I \in \mathcal{I}\},$ where $ \epsilon_{i,n}$ is the $n$-length vector whose $i^{th}$ component is 1 and all other components are zeros.

A matroid $\mathbb{M}$ is said to be representable over $\mathbb{F}_q$ if there exists one-dimensional vector subspaces $V_1,V_2, \dotso V_n$ of a vector space $E$ such that $\dim(\sum_{i \in X} V_i)=r(X), \forall X \subseteq \lceil n \rfloor$ and the set of vector subspaces $V_i, i \in \lceil n  \rfloor,$ is said to form a representation of $\mathbb{M}.$ The one-dimensional vector subspaces $V_i, i \in \lceil n \rfloor,$ can be described by a matrix $A$ over $\mathbb{F}_q$ with $n$ columns whose $i^{\text{th}}$ column spans $V_i.$ 
It is clear that the set of vector subspaces $V_i, i \in \lceil n  \rfloor,$ forms a representation of $\mathbb{M}$ if and only if it forms a representation of $\mathbb{D}(\mathbb{M}).$
 

\begin{example}
Consider the uniform matroid $U_{2,4}$ on the ground set $\lceil 4 \rfloor$ with the rank function given by,
\begin{displaymath}
   r(X)= \left\{
     \begin{array}{lr}
       \vert X \vert & :\text{if\;} \vert X \vert \leq 2\\
       2 & : \text{otherwise\;}
     \end{array}.
   \right.
\end{displaymath} 
Let $A=\begin{bmatrix} 1 & 0 & 1 & 1\\0 &1 &1 &2 \end{bmatrix}$ be a matrix over $\mathbb{F}_3.$ Let $V_i, i \in \lceil 4 \rfloor,$ denote the span of $i^{\text{th}}$ column of $A$ over $\mathbb{F}_3.$ It can be verified that the vector subspaces $V_1,V_2,V_3$ and $V_4$ form a representation of $U_{2,4}$ over $\mathbb{F}_3.$
\endproof
\end{example}

The notion of multi-linear representation of matroids was introduced in \cite{SiAs,Ma}, where it was shown that the non-Pappus matroid which is not representable over any field \cite{Ox}, has a multi-linear representation of dimension 2 over $\mathbb{F}_3$.
\begin{definition}[\cite{Ma}]
A matroid $\mathbb{M}=(\lceil n \rfloor,\rho)$ is said to be multi-linearly representable of dimension $k$ over $\mathbb{F}_q$ if there exists vector subspaces $V_1,V_2,\dotso,V_n$ of a vector space $E$ over $\mathbb{F}_q$ such that $dim(\sum_{i \in X} V_i)=kr(X), \forall X \subseteq \lceil n \rfloor.$ The vector subspaces $V_i,i\in\lceil n \rfloor,$ are said to form a multi-linear representation of dimension $k$ over $\mathbb{F}_q$ of the matroid $\mathbb{M}.$
\end{definition}

\begin{example}[\cite{SiAs}]
Consider the non-Pappus matroid whose geometric representation is shown in Fig. \ref{fig:matroid_non_pappus}. The rank function $r$ of the non-Pappus matroid can be described as follows: all subsets $X$ of $\lceil 9 \rfloor$ of cardinality less than or equal to two have rank equal to $\vert X \vert$ and those whose cardinality is greater than or equal to four have rank 3. Among those subsets with cardinality 3, if all its elements line on a line in the geometric representation shown in Fig. \ref{fig:matroid_non_pappus}, the rank is two; otherwise the rank is three. Let 

{\scriptsize $$A= \underbrace{\hspace{-.2 cm}\left[\begin{matrix} 1&\hspace{-.2 cm}0\\ 0&\hspace{-.2 cm}1\\0&\hspace{-.2 cm}0\\0&\hspace{-.2 cm}0\\0&\hspace{-.2 cm}0\\0&\hspace{-.2 cm}0 \end{matrix}\right.}_{A_1} \quad \underbrace{\begin{matrix} 1&\hspace{-.2 cm}0\\ 0&\hspace{-.2 cm}1\\0&\hspace{-.2 cm}0\\0&\hspace{-.2 cm}0\\1&\hspace{-.2 cm}0\\0&\hspace{-.2 cm}1 \end{matrix}}_{A_2} \quad \underbrace{\begin{matrix} 0&\hspace{-.2 cm}0\\ 0&\hspace{-.2 cm}0\\0&\hspace{-.2 cm}0\\0&\hspace{-.2 cm}0\\1&\hspace{-.2 cm}0\\0&\hspace{-.2 cm}1 \end{matrix}}_{A_3} \quad \underbrace{\begin{matrix} 1&\hspace{-.2 cm}0\\ 0&\hspace{-.2 cm}1\\1&\hspace{-.2 cm}0\\0&\hspace{-.2 cm}2\\0&\hspace{-.2 cm}1\\2&\hspace{-.2 cm}1 \end{matrix}}_{A_4} \quad \underbrace{\begin{matrix} 0&\hspace{-.2 cm}0\\ 0&\hspace{-.2 cm}0\\1&\hspace{-.2 cm}0\\0&\hspace{-.2 cm}1\\0&\hspace{-.2 cm}0\\0&\hspace{-.2 cm}0 \end{matrix}}_{A_5} \quad \underbrace{\begin{matrix} 1&\hspace{-.2 cm}0\\ 0&\hspace{-.2 cm}1\\2&\hspace{-.2 cm}1\\2&\hspace{-.2 cm}0\\0&\hspace{-.2 cm}1\\2&\hspace{-.2 cm}1 \end{matrix}}_{A_6} \quad \underbrace{\begin{matrix} 1&\hspace{-.2 cm}0\\ 0&\hspace{-.2 cm}1\\0&\hspace{-.2 cm}1\\1&\hspace{-.2 cm}2\\0&\hspace{-.2 cm}0\\0&\hspace{-.2 cm}0 \end{matrix}}_{A_7} \quad \underbrace{\begin{matrix} 1&\hspace{-.2 cm}0\\ 0&\hspace{-.2 cm}1\\1&\hspace{-.2 cm}0\\0&\hspace{-.2 cm}2\\1&\hspace{-.2 cm}1\\1&\hspace{-.2 cm}0 \end{matrix}}_{A_8}\quad \underbrace{\left.\begin{matrix} 0&\hspace{-.2 cm}0\\ 0&\hspace{-.2 cm}0\\1&\hspace{-.2 cm}0\\0&\hspace{-.2 cm}1\\1&\hspace{-.2 cm}0\\0&\hspace{-.2 cm}1 \end{matrix}\right]}_{A_9}.$$} 

\noindent be a matrix over $\mathbb{F}_3.$ The vector spaces given by the column span of the matrices $A_i, i \in \lceil 9 \rfloor,$ form a representation of the non-Pappus matroid.\endproof
\begin{figure}[htbp]
\centering
\includegraphics[totalheight=1.5 in,width=2.5in]{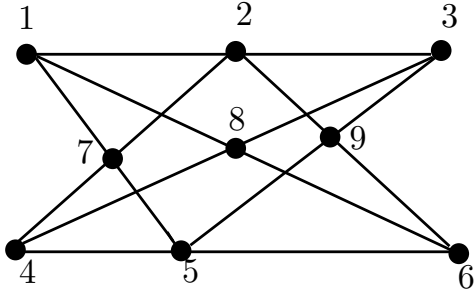}
\caption{The non-Pappus matroid}
\label{fig:matroid_non_pappus}
\end{figure}
\end{example}

Let $\mathbb{M}$ be a matroid with rank function $r.$ Define $\rho(X)=k r(X),X \subseteq \lceil n \rfloor,$ where $k$ is a positive integer. It can be verified that $\rho$ satisfies (D1)--(D3) and hence is the rank function of a discrete polymatroid. It is easy to verify that the vector subspaces over $V_1,V_2,\dotso,V_n$ over $\mathbb{F}_q$ form a representation of the discrete polymatroid with rank function $\rho$ if and only if they form a multi-linear representation of dimension $k$ over $\mathbb{F}_q$ of the matroid with rank function $r.$
\begin{example}
Continuing with Example 5, it can be easily shown that the matroid $U_{2,4}$ with rank function $r$ does not have a representation over $\mathbb{F}_2.$ But for the discrete polymatroid with rank function $\rho(X)=2r(X),$  the vector subspaces $V_i,i\in\lceil 4 \rfloor$ given in Example 4 form a representation over $\mathbb{F}_2$. Hence, the vector subspaces $V_i,i\in\lceil 4 \rfloor,$ given in Example 4 form a multi-linear representation of dimension 2 for the matroid $U_{2,4}$ over $\mathbb{F}_2.$ \endproof  
\end{example}
\section{Vector Linear Solvability of Networks and Representation of Discrete Polymatroids}
For a discrete polymatroid $\mathbb{D},$ let $\rho_{max}(\mathbb{D})=\max_{i \in \lceil n \rfloor} \rho(\lbrace i \rbrace).$

We define a discrete polymatroidal network as follows:
\begin{definition}
A network is said to be discrete polymatroidal with respect to a discrete polymatroid $\mathbb{D},$ if there exists a map $f: \mathcal{E} \rightarrow \lceil n \rfloor$ which satisfies the following conditions:
\begin{description}
\item [(DN1):]
$f$ is one-one on the elements of $\mathcal{S}.$
\item [(DN2):]
$\sum_{i \in f(\mathcal{S})} \rho_{max}(\mathbb{D}) \epsilon_{i}\in \mathbb{D}.$
\item [(DN3):]
$\rho(f(In(x)))=\rho(f(In(x) \cup Out(x))), \forall x \in \mathcal{V}.$
\end{description}
\end{definition}

The notion of a matroidal network was introduced in \cite{DoFrZe}. It can be verified that a network is matroidal with respect to a matroid $\mathbb{M}$ if and only if it is discrete polymatroidal with respect to $\mathbb{D}(\mathbb{M}).$

From the results in \cite{DoFrZe} and \cite{KiMe}, it follows that a network has scalar linear solution over $\mathbb{F}_q$ if and only if the network is matroidal with respect to a matroid representable over $\mathbb{F}_q.$ In the following theorem, we provide a generalization of this result for vector linear solvable networks, in terms of the representability of discrete polymatroids. 
\begin{figure*}[t]
\tiny
\begin{align}
\label{matrix_M_network_1}
&\hspace{3 cm}\underbrace{\hspace{-.01 cm}\left[\begin{matrix} 1&\hspace{-.2 cm}0\\ 0&\hspace{-.2 cm}1\\0&\hspace{-.2 cm}0\\0&\hspace{-.2 cm}0\\0&\hspace{-.2 cm}0\\0&\hspace{-.2 cm}0\\0&\hspace{-.2 cm}0 \\0&\hspace{-.2 cm}0\end{matrix}\right.}_{A_1} \quad\quad \underbrace{\begin{matrix} 0&\hspace{-.2 cm}0\\ 0&\hspace{-.2 cm}0\\1&\hspace{-.2 cm}0\\0&\hspace{-.2 cm}1\\0&\hspace{-.2 cm}0\\0&\hspace{-.2 cm}0\\0&\hspace{-.2 cm}0\\0&\hspace{-.2 cm}0 \end{matrix}}_{A_2} \quad\quad \underbrace{\begin{matrix} 0&\hspace{-.2 cm}0\\ 0&\hspace{-.2 cm}0\\0&\hspace{-.2 cm}0\\0&\hspace{-.2 cm}0\\1&\hspace{-.2 cm}0\\0&\hspace{-.2 cm}1\\0&\hspace{-.2 cm}0\\0&\hspace{-.2 cm}0 \end{matrix}}_{A_3} \quad\quad \underbrace{\begin{matrix} 0&\hspace{-.2 cm}0\\ 0&\hspace{-.2 cm}0\\0&\hspace{-.2 cm}0\\0&\hspace{-.2 cm}0\\0&\hspace{-.2 cm}0\\0&\hspace{-.2 cm}0\\1&\hspace{-.2 cm}0\\0&\hspace{-.2 cm}1 \end{matrix}}_{A_4} \quad\quad \underbrace{\begin{matrix} 1&\hspace{-.2 cm}0\\ 0&\hspace{-.2 cm}0\\0&\hspace{-.2 cm}0\\0&\hspace{-.2 cm}1\\0&\hspace{-.2 cm}0\\0&\hspace{-.2 cm}0\\0&\hspace{-.2 cm}0\\0&\hspace{-.2 cm}0 \end{matrix}}_{A_5} \quad\quad \underbrace{\begin{matrix} 0&\hspace{-.2 cm}0\\ 1&\hspace{-.2 cm}0\\0&\hspace{-.2 cm}1\\0&\hspace{-.2 cm}0\\0&\hspace{-.2 cm}0\\0&\hspace{-.2 cm}0\\0&\hspace{-.2 cm}0\\0&\hspace{-.2 cm}0 \end{matrix}}_{A_6} \quad\quad \underbrace{\begin{matrix} 0&\hspace{-.2 cm}0\\ 0&\hspace{-.2 cm}0\\0&\hspace{-.2 cm}0\\0&\hspace{-.2 cm}0\\1&\hspace{-.2 cm}0\\0&\hspace{-.2 cm}0 \\0&\hspace{-.2 cm}0\\0&\hspace{-.2 cm}1\end{matrix}}_{A_7} \quad\quad \underbrace{\begin{matrix} 0&\hspace{-.2 cm}0\\ 0&\hspace{-.2 cm}0\\0&\hspace{-.2 cm}0\\0&\hspace{-.2 cm}0\\0&\hspace{-.2 cm}0\\1&\hspace{-.2 cm}0\\0&\hspace{-.2 cm}1\\0&\hspace{-.2 cm}0 \end{matrix}}_{A_8}\quad\quad \underbrace{\begin{matrix} 0&\hspace{-.2 cm}0\\ 1&\hspace{-.2 cm}0\\0&\hspace{-.2 cm}0\\0&\hspace{-.2 cm}0\\0&\hspace{-.2 cm}1\\0&\hspace{-.2 cm}0\\0&\hspace{-.2 cm}0\\0&\hspace{-.2 cm}0 \end{matrix}}_{A_9}\quad\quad \underbrace{\begin{matrix} 0&\hspace{-.2 cm}0\\ 1&\hspace{-.2 cm}0\\0&\hspace{-.2 cm}0\\0&\hspace{-.2 cm}0\\0&\hspace{-.2 cm}0\\0&\hspace{-.2 cm}0\\0&\hspace{-.2 cm}0\\0&\hspace{-.2 cm}1 \end{matrix}}_{A_{10}}\quad\quad \underbrace{\begin{matrix} 0&\hspace{-.2 cm}0\\ 0&\hspace{-.2 cm}0\\1&\hspace{-.2 cm}0\\0&\hspace{-.2 cm}0\\0&\hspace{-.2 cm}1\\0&\hspace{-.2 cm}0\\0&\hspace{-.2 cm}0\\0&\hspace{-.2 cm}0 \end{matrix}}_{A_{11}}\quad\quad \underbrace{\left.\begin{matrix} 0&\hspace{-.2 cm}0\\ 0&\hspace{-.2 cm}0\\1&\hspace{-.2 cm}0\\0&\hspace{-.2 cm}0\\0&\hspace{-.2 cm}0\\0&\hspace{-.2 cm}0\\0&\hspace{-.2 cm}0\\0&\hspace{-.2 cm}1 \end{matrix}\right]}_{A_{12}}\\
 \hline
\label{matrix_M_network_2}
&\underbrace{\hspace{-.01 cm}\left[\begin{matrix} 1&\hspace{-.2 cm}0\\ 0&\hspace{-.2 cm}1\\0&\hspace{-.2 cm}0\\0&\hspace{-.2 cm}0\\0&\hspace{-.2 cm}0\\0&\hspace{-.2 cm}0\\0&\hspace{-.2 cm}0 \\0&\hspace{-.2 cm}0\end{matrix}\right.}_{A'_1} \quad\quad \underbrace{\begin{matrix} 0&\hspace{-.2 cm}0\\ 0&\hspace{-.2 cm}0\\1&\hspace{-.2 cm}0\\0&\hspace{-.2 cm}1\\0&\hspace{-.2 cm}0\\0&\hspace{-.2 cm}0\\0&\hspace{-.2 cm}0\\0&\hspace{-.2 cm}0 \end{matrix}}_{A'_2} \quad\quad \underbrace{\begin{matrix} 0&\hspace{-.2 cm}0\\ 0&\hspace{-.2 cm}0\\0&\hspace{-.2 cm}0\\0&\hspace{-.2 cm}0\\1&\hspace{-.2 cm}0\\0&\hspace{-.2 cm}1\\0&\hspace{-.2 cm}0\\0&\hspace{-.2 cm}0 \end{matrix}}_{A'_3} \quad\quad \underbrace{\begin{matrix} 0&\hspace{-.2 cm}0\\ 0&\hspace{-.2 cm}0\\0&\hspace{-.2 cm}0\\0&\hspace{-.2 cm}0\\0&\hspace{-.2 cm}0\\0&\hspace{-.2 cm}0\\1&\hspace{-.2 cm}0\\0&\hspace{-.2 cm}1 \end{matrix}}_{A'_4} \quad\quad \underbrace{\begin{matrix} 1&\hspace{-.2 cm}0\\ 0&\hspace{-.2 cm}0\\0&\hspace{-.2 cm}0\\0&\hspace{-.2 cm}1\\0&\hspace{-.2 cm}0\\0&\hspace{-.2 cm}0\\0&\hspace{-.2 cm}0\\0&\hspace{-.2 cm}0 \end{matrix}}_{A'_5} \quad\quad \underbrace{\begin{matrix} 0&\hspace{-.2 cm}0\\ 1&\hspace{-.2 cm}0\\0&\hspace{-.2 cm}1\\0&\hspace{-.2 cm}0\\0&\hspace{-.2 cm}0\\0&\hspace{-.2 cm}0\\0&\hspace{-.2 cm}0\\0&\hspace{-.2 cm}0 \end{matrix}}_{A'_6} \quad\quad \underbrace{\begin{matrix} 0&\hspace{-.2 cm}0\\ 0&\hspace{-.2 cm}0\\0&\hspace{-.2 cm}0\\0&\hspace{-.2 cm}0\\1&\hspace{-.2 cm}0\\0&\hspace{-.2 cm}0 \\0&\hspace{-.2 cm}0\\0&\hspace{-.2 cm}1\end{matrix}}_{A'_7} \quad\quad \underbrace{\begin{matrix} 0&\hspace{-.2 cm}0\\ 0&\hspace{-.2 cm}0\\0&\hspace{-.2 cm}0\\0&\hspace{-.2 cm}0\\0&\hspace{-.2 cm}0\\1&\hspace{-.2 cm}0\\0&\hspace{-.2 cm}1\\0&\hspace{-.2 cm}0 \end{matrix}}_{A'_8}\quad\quad \underbrace{\begin{matrix} 0&\hspace{-.2 cm}0\\ 1&\hspace{-.2 cm}0\\0&\hspace{-.2 cm}0\\0&\hspace{-.2 cm}0\\0&\hspace{-.2 cm}1\\0&\hspace{-.2 cm}0\\0&\hspace{-.2 cm}0\\0&\hspace{-.2 cm}0 \end{matrix}}_{A'_9}\quad\quad \underbrace{\begin{matrix} 0&\hspace{-.2 cm}0\\ 1&\hspace{-.2 cm}0\\0&\hspace{-.2 cm}0\\0&\hspace{-.2 cm}0\\0&\hspace{-.2 cm}0\\0&\hspace{-.2 cm}0\\0&\hspace{-.2 cm}0\\0&\hspace{-.2 cm}1 \end{matrix}}_{A'_{10}}\quad\quad \underbrace{\begin{matrix} 0&\hspace{-.2 cm}0\\ 0&\hspace{-.2 cm}0\\1&\hspace{-.2 cm}0\\0&\hspace{-.2 cm}0\\0&\hspace{-.2 cm}1\\0&\hspace{-.2 cm}0\\0&\hspace{-.2 cm}0\\0&\hspace{-.2 cm}0 \end{matrix}}_{A'_{11}}\quad\quad \underbrace{\begin{matrix} 0&\hspace{-.2 cm}0\\ 0&\hspace{-.2 cm}0\\1&\hspace{-.2 cm}0\\0&\hspace{-.2 cm}0\\0&\hspace{-.2 cm}0\\0&\hspace{-.2 cm}0\\0&\hspace{-.2 cm}0\\0&\hspace{-.2 cm}1 \end{matrix}}_{A'_{12}}\quad\quad
 \underbrace{\begin{matrix} 1&\hspace{-.2 cm}0\\ 0&\hspace{-.2 cm}0\\0&\hspace{-.2 cm}0\\0&\hspace{-.2 cm}0\\0&\hspace{-.2 cm}0\\0&\hspace{-.2 cm}0\\0&\hspace{-.2 cm}0 \\0&\hspace{-.2 cm}0\end{matrix}}_{A'_{13}} \quad\quad \underbrace{\begin{matrix} 1&\hspace{-.2 cm}0\\ 0&\hspace{-.2 cm}0\\0&\hspace{-.2 cm}0\\0&\hspace{-.2 cm}0\\0&\hspace{-.2 cm}0\\0&\hspace{-.2 cm}0\\0&\hspace{-.2 cm}0\\0&\hspace{-.2 cm}0 \end{matrix}}_{A'_{14}} \quad\quad \underbrace{\begin{matrix} 0&\hspace{-.2 cm}0\\ 0&\hspace{-.2 cm}0\\0&\hspace{-.2 cm}0\\1&\hspace{-.2 cm}0\\0&\hspace{-.2 cm}0\\0&\hspace{-.2 cm}0\\0&\hspace{-.2 cm}0\\0&\hspace{-.2 cm}0 \end{matrix}}_{A'_{15}} \quad\quad \underbrace{\begin{matrix} 0&\hspace{-.2 cm}0\\ 0&\hspace{-.2 cm}0\\0&\hspace{-.2 cm}0\\1&\hspace{-.2 cm}0\\0&\hspace{-.2 cm}0\\0&\hspace{-.2 cm}0\\0&\hspace{-.2 cm}0\\0&\hspace{-.2 cm}0 \end{matrix}}_{A'_{16}} \quad\quad \underbrace{\begin{matrix} 0&\hspace{-.2 cm}0\\ 0&\hspace{-.2 cm}0\\0&\hspace{-.2 cm}0\\0&\hspace{-.2 cm}0\\0&\hspace{-.2 cm}0\\1&\hspace{-.2 cm}0\\0&\hspace{-.2 cm}0\\0&\hspace{-.2 cm}0 \end{matrix}}_{A'_{17}} \quad\quad \underbrace{\begin{matrix} 0&\hspace{-.2 cm}0\\ 0&\hspace{-.2 cm}0\\0&\hspace{-.2 cm}0\\0&\hspace{-.2 cm}0\\0&\hspace{-.2 cm}0\\0&\hspace{-.2 cm}0\\1&\hspace{-.2 cm}0\\0&\hspace{-.2 cm}0 \end{matrix}}_{A'_{18}} \quad\quad \underbrace{\begin{matrix} 0&\hspace{-.2 cm}0\\ 0&\hspace{-.2 cm}0\\0&\hspace{-.2 cm}0\\0&\hspace{-.2 cm}0\\0&\hspace{-.2 cm}0\\1&\hspace{-.2 cm}0 \\0&\hspace{-.2 cm}0\\0&\hspace{-.2 cm}0\end{matrix}}_{A'_{19}} \quad\quad \underbrace{\left.\begin{matrix} 0&\hspace{-.2 cm}0\\ 0&\hspace{-.2 cm}0\\0&\hspace{-.2 cm}0\\0&\hspace{-.2 cm}0\\0&\hspace{-.2 cm}0\\0&\hspace{-.2 cm}0\\1&\hspace{-.2 cm}0\\0&\hspace{-.2 cm}0 \end{matrix}\right]}_{A'_{20}}
\end{align}
\hrule
\end{figure*}
\begin{theorem}
A network has a $k$ dimensional vector linear solution over $\mathbb{F}_q$ if and only if it is discrete polymatroidal with respect to a discrete polymatroid  $\mathbb{D}$ representable over $\mathbb{F}_q$ with $\rho_{max}(\mathbb{D})=k.$ 
\begin{proof}
Assume the edge set $\mathcal{E}$ to be $\lceil l \rfloor,$ with the set of input edges $\mathcal{S}=\lceil m \rfloor.$ Assume the set of intermediate edges to be $\{m+1,m+2,\dotso, l\rbrace,$ with the edges in the set arranged in the ancestral ordering, which exists since the networks considered in the paper are acyclic.

First we prove the if part. Consider a network which is discrete polymatroidal with a respect to a representable discrete polymatroid $\mathbb{D}(V_1,V_2,\dotso,V_n)$ which is denoted as $\mathbb{D}$ for brevity, with $\rho_{max}(\mathbb{D})=\max_{i \in \lceil n \rfloor}dim(V_i)=k.$ Let $f$ be the mapping from edge set of the network $\mathcal{E}$ to the ground set $\lceil n \rfloor$ of the discrete polymatroid  which satisfies (DN1)-(DN3).  Since the map $f$ is one-one on the elements of $\mathcal{S},$ assume $f(\mathcal{S})=\lceil m \rfloor.$ Let $v=\sum_{i \in \lceil m \rfloor} \epsilon_{i,n},$ where $\epsilon_{i,n}$ is the vector of length $n$ whose $i^{\text{th}}$ component is one and all other components are zeros. From (DN2) it follows that,
\begin{equation}
\label{proof_eqn1}
k \vert v(A) \vert \leq dim \left(\sum_{s \in A} V_{s}\right), \forall A \subseteq \lceil n \rfloor. 
\end{equation}

It is claimed that without loss of generality, we can take $\lceil n \rfloor$ to be the image of $f.$ Otherwise, let the image of $f$ be the set $\{i_1,i_2,\dotso,i_t \}\subset \lceil n \rfloor.$ We show that the network is discrete polymatroidal with respect to the discrete polymatroid $\mathbb{D}'=\mathbb{D}(V_{i_1},V_{i_2},\dotso V_{i_t}),$ with $f$ as the network discrete polymatroid mapping. (DN1) and (DN3) follow from the fact that the network is discrete polymatroidal with respect to $\mathbb{D}$ with $f$ as the network discrete polymatroid mapping. Let $u=k \sum_{i \in \lceil m \rfloor} \epsilon_{i,t},$ where $\epsilon_{i,t}$ is the vector of length $t$ whose $i^{\text{th}}$ component is one and all other components are zeros. To prove that (DN2) is also satisfied, it needs to be shown that $k\vert u(A') \vert \leq dim(\sum_{r \in A'} V_{i_r}), \forall A' \subseteq \lceil t \rfloor,$ which follows from $\eqref{proof_eqn1}$ and from the facts that $\{i_r : r \in A'\} \subseteq \lceil n \rfloor$ and $u(A')=v(\{i_r : r \in A'\}).$ 

It is claimed that $dim(\sum_{i \in \lceil n \rfloor}V_i)=k m.$ The proof of the claim is as follows: Define $s_0=\lceil m \rfloor.$ Let $s_1=s_0 \cup \lbrace f(m+1)\rbrace.$ Since the edges in the set $\{m+1,m+2, \dotso, l \}$ are arranged in ancestral ordering, $In(head(m+1))$ is contained in $s_0.$ Hence, from (DN3) we have $\rho(s_1)=dim(\sum_{i \in s_0} V_i + V_{f(m+1)})=  dim(\sum_{i \in s_0}V_i)=\rho(s_0).$ Iteratively, defining $s_{i+1}=s_{i} \cup \lbrace f(m+i+1)\rbrace,$ using a similar argument, we have $\rho(s_{i+1})=\rho(s_0).$ Hence, we have $\rho(s_{l-m})=\rho(s_0)=\rho(\lceil m \rfloor).$ But $s_{l-m}=\lceil n \rfloor,$ since the image of $f$ is $\lceil n \rfloor.$ Hence, we have, $\rho(\lceil n \rfloor)=\rho(\lceil m \rfloor).$  Since the network is discrete polymatroidal, $\sum_{i \in \lceil m \rfloor} k \epsilon_{i} \in \mathbb{D}$ and from the definition of $\mathbb{D},$ it follows that $k m \leq dim \left(\sum_{i \in \lceil m \rfloor} V_i \right).$ But from (D2) we have, $dim \left(\sum_{i \in \lceil m \rfloor} V_i \right)=\rho(\lceil m \rfloor)\leq \rho(\lbrace 1\rbrace)+ \rho(\lbrace 2,3,\dotso,m \rbrace)\leq  \sum_{i \in \lceil 2 \rfloor} \rho(\lbrace i\rbrace)+\rho(\lbrace 3,4,\dotso, m \rbrace) \leq\dotso \leq \sum_{i \in \lceil m \rfloor} \rho(\lbrace i \rbrace)\leq k m.$ Hence, $\rho(\lceil n \rfloor)=\rho(\lceil m \rfloor)=k m.$ Hence $dim(\sum_{i \in \lceil n \rfloor}V_i)=k m.$ 

The vector subspace $V_i, i \in \lceil n \rfloor,$ can be written as the column span of a matrix $A_i$ of size $km \times k.$  Also, since $dim(\sum_{i \in \lceil m \rfloor}V_i)=k m,$ the matrix $B=[A_1 \;A_2 \dotso\;A_m]$ is invertible and hence can be taken to be the $ km \times km$ identity matrix (Otherwise, it is possible to choose $A'_i=B^{-1}A_i,$ and $V'_i$ to be the column span of $A'_i,$ so that $\mathbb{D}(V'_1,V'_2,\dotso,V'_n)=\mathbb{D}(V_1,V_2,\dotso,V_n)$ and $[A'_1,A'_2,\dotso,A'_m]$ is the identity matrix).

Taking the global encoding matrices $M_i$ to be $A_{f(i)},$ we get a $k$-dimensional network coding solution over $\mathbb{F}_q$ for the network. Since $[A_1 \;A_2 \dotso\;A_m]$ is the identity matrix, $f_i(x)=x A_i=x_i, \forall i \in \mathcal{S}$ and (N1) is satisfied. For a vertex $v \in \mathcal{V},$ with $In(v)=\lbrace i_1,i_2,\dotso,i_t \rbrace,$ from (DN3) it follows that, $rank([A_{f(i_1)}\;A_{f(i_2)}\dotso\;A_{f(i_t)}\;A_{f(j)}])=rank([A_{f(i_1)}\;A_{f(i_2)}\dotso\;A_{f(i_t)}]),$ for all $j \in Out(v).$ Hence, the matrix $A_{f(j)}$ can be written as $\sum_{p=1}^{t}  A_{f(i_p)} W_p,$ where $W_p \in \mathbb{F}_q^{k \times k},$  which shows that (N2) and (N3) are satisfied.  

To prove the only if part, consider a network which has a $k$-dimensional vector linear solution over $\mathbb{F}_q.$ Take the vector subspace $V_i$ to be the column span of the global encoding matrix $M_i, i \in \lceil l \rfloor.$ Consider the discrete polymatroid $\mathbb{D}(V_1,V_2,\dotso,V_l).$ The edge $i \in \mathcal{E}$ is mapped by the function $f$ to the element $i$ in the ground set $\lceil l \rfloor$ of the discrete polymatroid. It can be easily seen that (DN1)-(DN3) are satisfied and hence the network is discrete polymatroidal with respect to $\mathbb{D}(V_1,V_2,\dotso,V_l).$ Also, we have $\rho_{max}(\mathbb{D}(V_1,V_2,\dotso,V_l))=\max_{i \in \lceil l \rfloor} rank (V_i)=k.$ 
\end{proof}
\end{theorem}

It is important to note that the discrete polymatroid $\mathbb{D}$ in Theorem 1 need not be unique.  A network can admit more than one $k$ dimensional vector linear solution over $\mathbb{F}_q$ and from these solutions it may be possible to obtain multiple discrete polymatroids with respect to which the network is discrete polymatroidal, as illustrated in the following example. 
\begin{example}
Consider the M-network shown in Fig. \ref{fig:M_network}, introduced in \cite{MeEfHoKa}. It was shown in \cite{MeEfHoKa} that the M-network has a 2 dimensional vector linear solution which is in fact a vector routing solution, but it does not admit scalar linear solution over any field. It was shown in \cite{DoFrZe} that the M-network is not matroidal. But, from Theorem 1, it follows that the M-network is discrete polymatroidal with respect to a discrete polymatroid $\mathbb{D}$ which has $\rho_{max}(\mathbb{D})=2.$ We consider two possible solutions for the M-network, from which it is possible to obtain two different discrete polymatroids with respect to which the M-network is discrete polymatroidal.\\ 
\underline{\textit{Solution 1:}}
 Assume the global encoding matrices of edge $i,i\in \lceil 12 \rfloor,$ to be the matrix $A_i$ given in \eqref{matrix_M_network_1} at the top of this page. Take $A_5$ to be the global encoding matrix of the  edges $13,14,15,16$ and  $A_8$ to be that of $17,18,19,20.$ The solution thus obtained for the M-network is as shown in Fig. \ref{fig:sol1}. Let the network discrete polymatroid mapping $f_1$ be defined as follows: 
\begin{displaymath}
   f_1(i)= \left\{
     \begin{array}{ll}
       &i  :i \in \{1,2,\dotso, 12\}\\
       &5  :i \in \{\:13,14,15,16\}\\
       &8 : i \in \{\:17,18,19,20\}
     \end{array}.
   \right.
\end{displaymath}  
Define $V_i$ to be the column span of $A_i,i\in\lceil 12 \rfloor.$ It can be verified that the M-network is discrete polymatroidal with respect to $\mathbb{D}(V_1,V_2,\dotso V_{12}),$ with $f_1$ being the network discrete polymatroid mapping. 

 It can be deduced from Definition 2 that the vector subspaces (excluding the trivial zero vector subspaces) which form a multi-linear representation of dimension $k$ for a matroid should be $k$-dimensional. Note that the vector subspaces $V_i,i \in \lceil 12 \rfloor,$ have dimension 2 and they from a representation for the discrete polymatroid $\mathbb{D}(V_1,V_2,\dotso V_{12}).$ Despite having their dimensions to be equal, the vector subspaces $V_i,i \in \lceil 12 \rfloor,$ cannot form a multi-linear representation of dimension 2 for any matroid. This follows from the fact that the M-network is not matroidal with respect to any matroid.\\
\underline{\textit{Solution 2:}}
 Assume the global encoding matrices of edge $i,i\in \lceil 20 \rfloor,$ to be the matrix $A'_i$ (defined in \eqref{matrix_M_network_2} at the top of this page). The solution thus obtained for the M-network is as shown in Fig. \ref{fig:sol2}. Let the network discrete polymatroid mapping $f_2(i)=i, i \in \lceil 20 \rfloor.$
Define $V'_i$ to be the column span of $A'_i,i\in\lceil 20 \rfloor.$ It can be verified that the M-network is discrete polymatroidal with respect to $\mathbb{D}(V'_1,V'_2,\dotso V'_{20}),$ with $f_2$ being the network discrete polymatroid mapping.

 Note that all the vector subspaces $V_i, i\in \lceil 12 \rfloor,$ in Solution 1 have the same dimension 2. In contrast, in Solution 2, the vector subspaces $V'_1,V'_2,\dotso, V'_{12}$ have dimension 2, while the vector subspaces $V'_{13},V'_{14},\dotso, V'_{20}$ have dimension 1. The M-network is discrete polymatroidal with respect to two different discrete polymatroids $\mathbb{D}(V_1,V_2,\dotso,V_{12})$ and $\mathbb{D}(V'_1,V'_2,\dotso,V'_{20}).$ \endproof
\begin{figure}[t]
\centering
\subfigure[Solution 1]{
\includegraphics[totalheight=3in,width=3in]{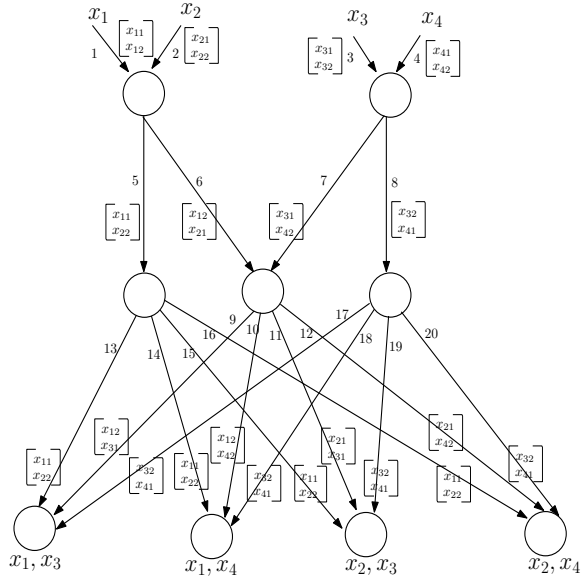}
\label{fig:sol1}
}
\subfigure[Solution 2]{
\includegraphics[totalheight=3in,width=3in]{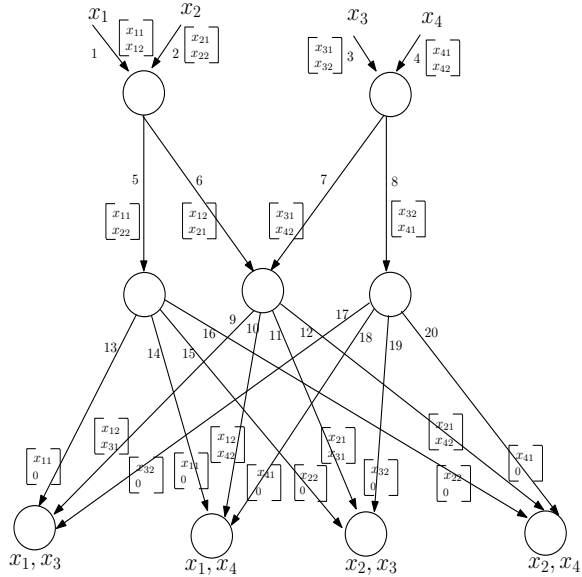}
\label{fig:sol2}
}
\caption{The M-network}
\label{fig:M_network}
\end{figure}
\end{example}
\section{Construction of Networks from Discrete Polymatroids}
In this section, an algorithm to construct a network from a discrete polymatroid $\mathbb{D}$ is provided. If the discrete polymatroid $\mathbb{D}$ is representable over $\mathbb{F}_q$ with $\rho_{max}(\mathbb{D})=k,$ the constructed network has a $k$-dimensional vector linear solution over $\mathbb{F}_q.$

Dougherty et. al. provided a construction procedure in \cite{DoFrZe} to obtain networks from a matroid, with the resulting network being scalar linearly solvable if the matroid is representable. The construction in \cite{DoFrZe} is heavily dependent on the set of circuits of the matroid from which the network is constructed. 

In Section II B, the connection between the independent sets of a matroid $\mathbb{M}$ and the vectors which belongs to the discrete polymatroid $\mathbb{D}(\mathbb{M})$ was discussed. 
 We provide some useful definitions for a discrete polymatroid $\mathbb{D},$ which when specialized to $\mathbb{D}(\mathbb{M})$ are related to the well known notions of dependent sets and circuits of $\mathbb{M}.$
\begin{definition}
For a discrete polymatroid $\mathbb{D},$ a vector $u \in \mathbb{Z}_{\geq 0}^{n}$ is said to be an excluded vector if the $i^{\text{th}}$ component of $u$ is    less than or equal to $\rho(\lbrace i \rbrace), \forall i \in \lceil n \rfloor,$ and $u \notin \mathbb{D}.$
\end{definition}
\begin{example}
For the discrete polymatroid considered in Example \ref{example_2d}, the excluded vectors are the points indicated by `x' in Fig. \ref{example_2d_dep}.
\begin{figure}[htbp]
\centering
\includegraphics[totalheight=2.7 in,width=3.6in]{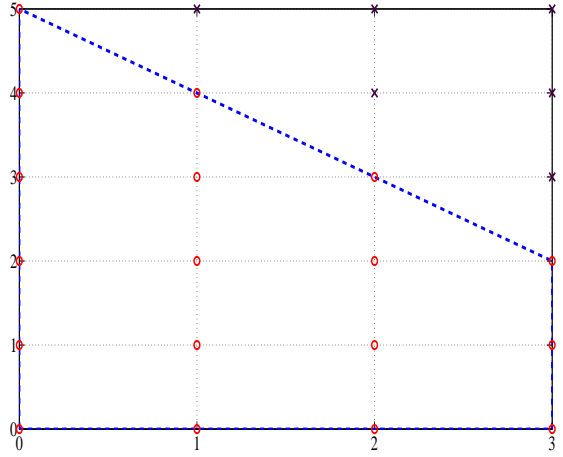}
\caption{Diagram showing the excluded vectors for the discrete polymatroid defined in Example \ref{example_2d}}
\label{example_2d_dep}
\end{figure}
\end{example}

For a discrete polymatroid $\mathbb{D},$ let $\mathcal{D}(\mathbb{D})$ denote the set of excluded vectors. 
For a vector $u \in \mathbb{Z}_{\geq 0}^{n},$ let $(u)_{>0}$ denotes the set of indices corresponding to the non-zero components of $u.$

For a matroid $\mathbb{M},$ the set of excluded vectors of the discrete polymatroid $\mathbb{D}(\mathbb{M})$ is in one-one correspondence with the set of dependent sets of $\mathbb{M},$ i.e., the set of dependent sets of $\mathbb{M}$ is given by $\{(u)_{>0}: u \in \mathcal{D}(\mathbb{D}(\mathbb{M}))\}.$

Let  $\mathcal{D}_i(\mathbb{D}), i \in \lceil n \rfloor$ denote the set of excluded vectors whose $i^{\text{th}}$ component is 1. 
For a matroid $\mathbb{M},$ the set $\mathcal{D}_i(\mathbb{D}(\mathbb{M}))$ uniquely identifies the set of dependent sets of $\mathbb{M}$ which contain the element $i,$ i.e., the set of dependent sets of $\mathbb{M}$ which contain $i$ is given by, $\{(u)_{>0} : u \in \mathcal{D}_i(\mathbb{D}(\mathbb{M}))\}.$

Let $\mathcal{C}_i(\mathbb{D}), i \in \lceil n \rfloor$ denote the set of vectors $u \in \mathcal{D}_i(\mathbb{D})$ which satisfy the following three conditions:
\begin{enumerate}
\item
$u-\epsilon_i \in \mathbb{D}.$
\item
There does not exist $ v \neq u \in \mathcal{D}_i(\mathbb{D})$ for which  $v < u.$ 
\item
$(v)_{>0} \not \subset (u)_{>0},$ for all $v \neq u \in \mathcal{D}_i(\mathbb{D}).$ 
\end{enumerate} 
For a matroid $\mathbb{M},$ the set $\mathcal{C}_i(\mathbb{D}(\mathbb{M}))$ uniquely determines the set of circuits which contain the element $i,$ i.e., the set of circuits which contain the element $i$ is given by, $\{(u)_{>0}, u \in \mathcal{C}_i(\mathbb{D}(\mathbb{M}))\}.$

Let $\mathcal{R}(\mathbb{D})$ denote the set of $v \in \mathbb{D}$ which satisfies the following two conditions:
\begin{enumerate}
\item
All the non-zero components of $v$ are equal to $\rho_{max}(\mathbb{D}).$
\item
Suppose $v' \neq v \in \mathbb{D}$ has all the non-zero components to be equal to $\rho_{max}(\mathbb{D}),$ $(v)_{>0} \not \subset (v')_{>0}.$
\end{enumerate}  
For a matroid $\mathbb{M},$ we have $\mathcal{R}(\mathbb{D}(\mathbb{M}))=\mathcal{B}(\mathbb{D}(\mathbb{M})).$
\begin{example}
For the discrete polymatroid considered in Example \ref{example1}, the set of vectors $\mathcal{D}_i(\mathbb{D}),i\in\lceil 4 \rfloor,$ are as given below:

{\footnotesize
\begin{align*}
&\mathcal{D}_1(\mathbb{D})=\lbrace (1,0,2,2),(1,1,1,2),(1,1,2,1),(1,1,2,2), (1,2,0,2),\\
&\hspace{1.5 cm}(1,2,1,1,),(1,2,1,2),(1,2,2,0),(1,2,2,1),(1,2,2,2) \rbrace,\\
&\mathcal{D}_2(\mathbb{D})=\lbrace (0,1,2,2),(1,1,1,2),(1,1,2,1),(1,1,2,2), (2,1,0,2),\\
&\hspace{1.5 cm}(2,1,1,1,),(2,1,1,2),(2,1,2,0),(2,1,2,1),(2,1,2,2) \rbrace,\\
&\mathcal{D}_3(\mathbb{D})=\lbrace     ( 0,     2  ,   1   ,  2),
                                        ( 1,     1,     1,     2),
                                         (1,     2 ,    1,     1),
     (1     ,2,     1     ,2),
     (2     ,0 ,    1    , 2),\\
    &\hspace{1.5 cm} (2    , 1  ,   1   ,  1),
     (2   ,  1   ,  1   ,  2),
     (2  ,   2    , 1  ,   0),
     (2 ,    2     ,1 ,    1),
     (2,     2     ,1,     2) \rbrace,\\
&\mathcal{D}_4(\mathbb{D})=\lbrace   (0,     2,     2,     1),
     (1   ,  1   ,  2     ,1),
     (1   ,  2   ,  1     ,1),
     (1   ,  2   ,  2     ,1),
     (2   ,  0   ,  2     ,1),\\
      &\hspace{1.5 cm} (2   ,  1   ,  1   ,  1),
     (2   ,  1   ,  2    , 1),
     (2   ,  2   ,  0  ,   1),
     (2   ,  2   ,  1 ,    1),
     (2   ,  2    , 2,     1)
 \rbrace.
\end{align*}}
The set of vectors $\mathcal{C}_i(\mathbb{D}),i \in \lceil 4 \rfloor$ are as given below.
\begin{align*}
\mathcal{C}_1(\mathbb{D})=\lbrace (1,0,2,2), (1,2,0,2),(1,2,2,0) \rbrace,\\
\mathcal{C}_2(\mathbb{D})=\lbrace (0,1,2,2), (2,1,0,2),(2,1,2,0) \rbrace,\\
\mathcal{C}_3(\mathbb{D})=\lbrace (2,2,1,0), (0,2,1,2),(2,0,1,2) \rbrace,\\
\mathcal{C}_4(\mathbb{D})=\lbrace (2,2,0,1), (0,2,2,1),(2,0,2,1) \rbrace.
\end{align*}
The set $\mathcal{R}(\mathbb{D})$ is given by,
\begin{align*}
\mathcal{R}(\mathbb{D})&=\lbrace (0,0,2,2),(0,2,0,2),(0,2,2,0),(2,0,0,2),\\
&\hspace{4.5 cm}(2,0,2,0),(2,2,0,0) \rbrace.
\end{align*}
\end{example}

Now we proceed to give the construction algorithm.\\

\underline{\textbf{{ALGORITHM 1}}}\\
\underline{\emph{Step 1:}}
From the set $\mathcal{R}(\mathbb{D}),$ choose a vector $v$ for which $\vert (v)_{>0} \vert$ is maximum. For every $i \in (v)_{>0},$ add a node $i$ to the network with an input edge $e_i$ which generates the message $x_i.$ Let $f(e_i)=i.$ Define $M=T=(v)_{>0}.$\\
\underline{\emph{Step 2:}}
For $i \in \lceil n \rfloor \notin T,$ find a vector $u \in \mathcal{C}_i(\mathbb{D}),$ for which \mbox{$(u-\epsilon_i)_{>0} \subseteq T.$} Add a new node $i'$ to the network with incoming edges from all the nodes which belong to $(u-\epsilon_i)_{>0}.$ Also, add a node $i$ with a single incoming edge from $i',$ denoted as $e_{i',i}.$ Define $f(e)=head(e), \forall e \in In(i)$ and $f(e_{i',i})=i.$ Let $T=T \cup \lbrace i\rbrace.$
Repeat step 2 until it is no longer possible.\\
\underline{\emph{Step 3:}} 
For $i \in M,$ choose a vector $u$ from $\mathcal{C}_i(\mathbb{D})$ for which $(u)_{>0} \subseteq T.$ Add a new node $n$ to the network which demands message $x_i$ and which has connections from the nodes in $(u-\epsilon_i)_{>0}.$ Define $f(e)=head(e), \forall e \in In(n).$  Repeat this step as many number of times as desired. 

The network constructed using ALGORITHM 1, is discrete polymatroidal with respect to $\mathbb{D}$ with the network discrete polymatroid mapping $f$ defined in the algorithm. Hence, if $\mathbb{D}$ is representable over $\mathbb{F}_q,$ then the constructed network admits a vector linear solution over $\mathbb{F}_q,$ as shown in the following theorem.
\begin{theorem}
A network constructed using ALGORITHM 1 from a discrete polymatroid $\mathbb{D}$  which is representable over $\mathbb{F}_q$ with $\rho_{max}(\mathbb{D})=k,$ admits a vector linear solution of dimension $k$ over $\mathbb{F}_q.$
\begin{proof}
The theorem is proved by showing that the constructed network is discrete polymatroidal with respect to $\mathbb{D}$ with the network discrete polymatroid mapping $f$ defined in ALGORITHM 1. Clearly, Step 1 of ALGORITHM 1 ensures that (DN1) and (DN2) are satisfied. 

The nodes in the network constructed using ALGORITHM 1 can be classified into four kinds: (i) nodes added in Step 1 which belong to the set $M,$ (ii) nodes added in step 2 which are labelled $i', i \in \lceil n \rfloor,$ (iii) nodes added in step 2 which are labelled $i, i \in \lceil n \rfloor$ and (iv) nodes added in Step 3 which demand messages. For a node $x$ of kind (i) or kind (iii), since the in-degree is one and all the outgoing edges are mapped by $f$ to the same element in $\lceil n \rfloor,$ $f(In(x))=f(In(x) \cup Out(x))$ and hence $\rho(f(In(x)))=\rho(f(In(x) \cup Out(x))).$

Consider a node $i' \in \lceil n \rfloor$ of kind (ii). Let $e_{i',i}$ denote the edge connecting $i'$ and $i.$ Let $u^i \in \mathcal{C}_i(\mathbb{D})$ denote the vector which was used in Step 2 while adding the node $i$ and $i'$ to the network. Since $f(e_{i',i})=i,$ we need to show that $\rho(f(In(i')))= \rho(f(In(i'))\cup \{i\}).$ Since $f(In(i'))=(u^i-\epsilon_i)_{>0}$ and $(u^i-\epsilon_i)_{>0} \cup \{i\}=(u^i)_{>0},$ it needs to be shown that $\rho\left(\left(u^i-\epsilon_i\right)_{>0}\right)=\rho\left((u^i)_{>0}\right),$ i.e., $dim\left(\sum_{j \in (u^i)_{>0}}V_j\right)=dim\left(\sum_{j \in \left(u^i-\epsilon_i\right)_{>0}}V_j\right).$ Let $a^i=(u^i-\epsilon_i).$ Since $a^i \in \mathbb{D},$ for all $A \subseteq \lceil n \rfloor,$ we have, 
\begin{equation}
\label{eqn2_thm}
\vert a^i(A)\vert\leq dim\left(\sum_{j \in A} V_j \right).
\end{equation}
Since $u^i \notin \mathbb{D},$ we have,
\begin{equation}
\label{eqn3_thm}
dim \left(\sum_{j \in A'} V_j\right)<\vert u^i(A') \vert,
\end{equation}
for some $A' \subseteq \lceil n \rfloor.$ Clearly $A'$ should contain $i,$ otherwise $\vert a^i(A')\vert=\vert u^i(A') \vert$ and, \eqref{eqn2_thm} and \eqref{eqn3_thm} cannot be simultaneously satisfied. Since $A'$ contains $i,$ we have $\vert u^i(A') \vert=\vert a^i(A')\vert+1.$ Hence, from \eqref{eqn2_thm} and \eqref{eqn3_thm} we get $dim\left(\sum_{j \in A'} V_j \right)= \vert a^i(A')\vert.$

\begin{figure}[htbp]
\centering
\includegraphics[totalheight=1.8 in,width=2in]{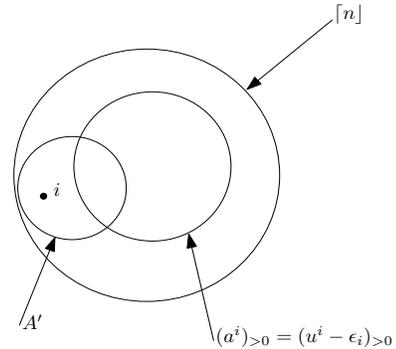}
\caption{Pictorial depiction of the sets $\lceil n \rfloor,$ $(a^i)_{>0}$ and $A'$ used in the proof of Theorem 2.}
\label{fig:figure_proof_theorem2}
\end{figure}
The sets $\lceil n \rfloor,$ $(a^i)_{>0}$ and the set $A'$ containing $i$ are pictorially depicted in Fig. \ref{fig:figure_proof_theorem2}. We have,
$\displaystyle{dim \left(\sum_{j \in (a^i)_{>0} \cap A'} V_j \right) \leq dim \left(\sum_{j \in A'} V_j \right)=\vert a^i(A')\vert.}$ Since \mbox{$a^i \in \mathbb{D},$} we have, $$\displaystyle{\vert a^i(A')\vert = \left\vert a^i\left(\left(a_i\right)_{>0} \cap A'\right)\right\vert \leq dim \left ( \sum_{j \in \left(a_i\right)_{>0}\cap A'} V_j \right).}$$ Hence, $dim \left ( \sum_{j \in \left(a_i\right)_{>0}\cap A'} V_j \right)=dim \left(\sum_{j \in A'} V_j \right).$ Since $i \in A',$ it follows that $dim \left( \sum_{j \in \left(a_i\right)_{>0}\cap A'} V_j +V_i\right)=dim\left(\sum_{j \in \left(a_i\right)_{>0}\cap A'} V_j \right).$ As a result, we have, 
{\scriptsize 
\begin{align*}
dim \left( \sum_{j \in \left(a_i\right)_{>0}\cap A'} V_j +V_i +\sum_{j \in \left(a_i\right)_{>0}\setminus A'} V_j\right)&\\
&\hspace{-3 cm}=dim\left(\sum_{j \in \left(a_i\right)_{>0}\cap A'} V_j +\sum_{j \in \left(a_i\right)_{>0}\setminus A'} V_j\right),
\end{align*}} 

\noindent i.e., $dim\left(\sum_{j \in (u^i)_{>0}}V_j\right)=dim\left(\sum_{j \in \left(u^i-\epsilon_i\right)_{>0}}V_j\right).$

Following a procedure exactly similar to the one used for a node kind (ii), it can be shown that $\rho(f(In(x)))=\rho(f(In(x) \cup Out(x)))$ for a node $x$ of kind (iv).  This completes the proof of Theorem 2.
\end{proof}
\end{theorem}
The construction procedure is illustrated using the following examples. 
\begin{example}
Continuing with Example 10, the construction procedure for the discrete polymatroid considered in Example \ref{example1} is described below.
\end{example}
\underline{\emph{Step 1:}}
For every vector $v\in \mathcal{R}(\mathbb{D}),$ $\vert (v)_{>0}\vert=2$ and hence any one of the vectors from $\mathcal{R}(\mathbb{D})$ can be chosen. Choose $v=(2,2,0,0)$ for which $(v)_{>0}=\lbrace 1,2 \rbrace.$ Add nodes 1 and 2 to the network with input edges generating the messages $x_1$ and $x_2$ respectively. We have $M=T=\{1,2\}.$
\begin{figure}[h]
\centering
\includegraphics[totalheight=.35 in,width=1in]{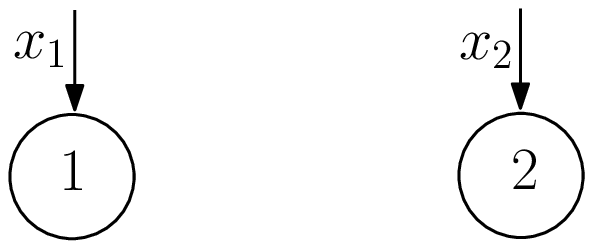}
\end{figure}\\
\underline{\emph{Step 2:}}
\begin{itemize}
\item
Pick $u=(2,2,1,0)$ from $\mathcal{C}_3(\mathbb{D}).$ Note that $(u-\epsilon_3)_{>0}=\{1,2\}\subseteq T.$ Add node $3'$ to the network with incoming edges from nodes $1$ and $2.$ Also, add node $3$ to the network which has its only incoming edge from $3'.$ $T=\lbrace 1,2,3\rbrace.$ 
\begin{figure}[h]
\centering
\includegraphics[totalheight=1 in,width=.8in]{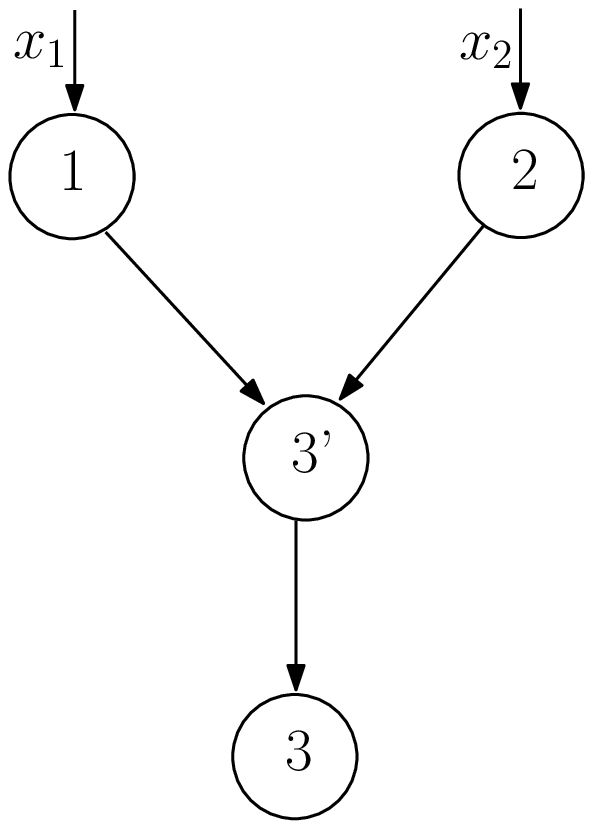}
\end{figure}
\item
Pick $u=(2,2,0,1) \in \mathcal{C}_4(\mathbb{D})$ for which $(u-\epsilon_4)_{>0}=\{1,2\}\subseteq T.$ Add node $4'$ to the network with incoming edges from nodes $1$ and $2.$ Also, add node $4$ with incoming edge from $4'.$ $T=\lbrace 1,2,3,4\rbrace.$ 
\begin{figure}[h]
\centering
\includegraphics[totalheight=2.5 in,width=1.25in]{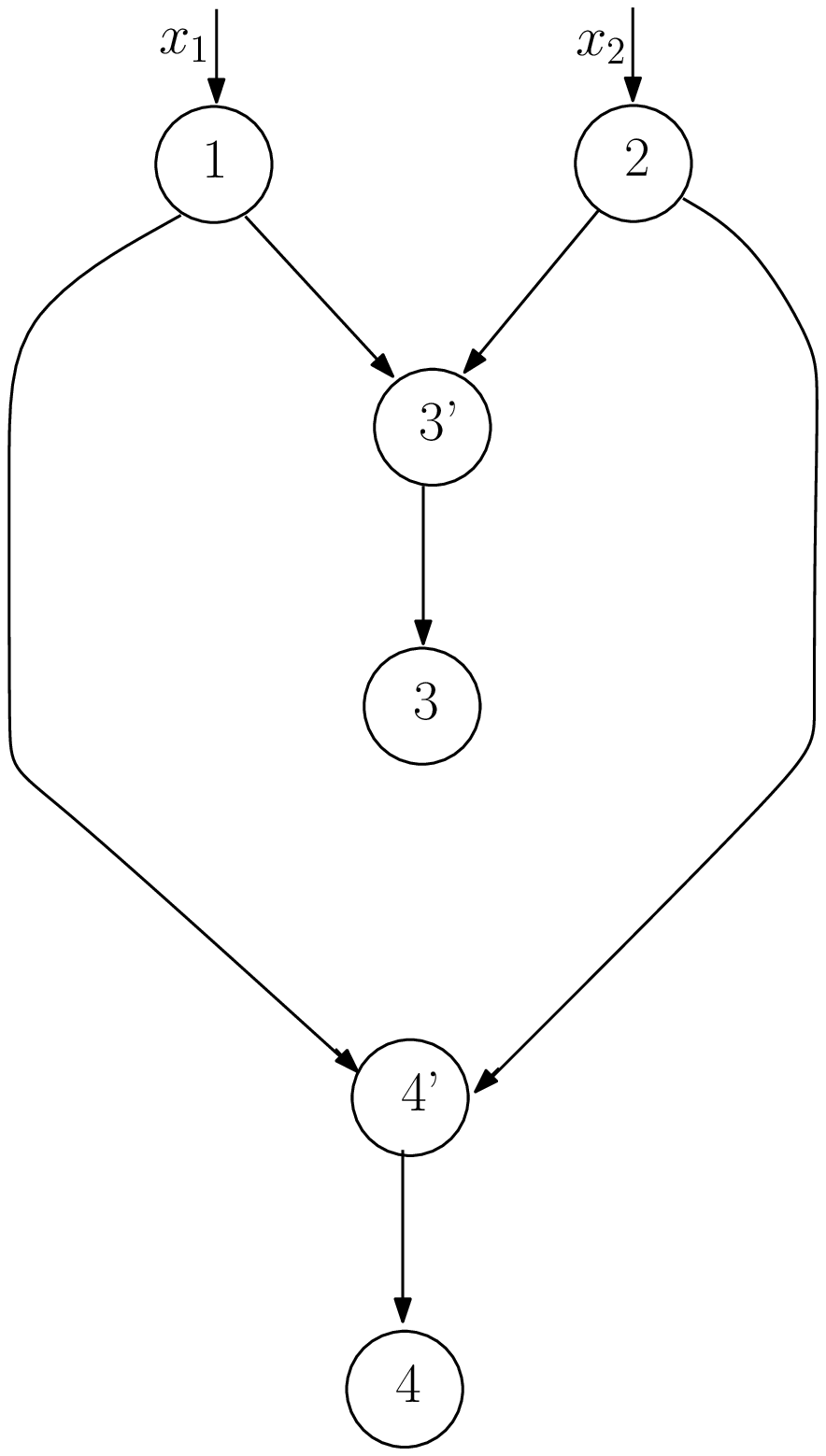}
\end{figure}
\end{itemize}
\underline{\emph{Step 3:}}
\begin{itemize}
\item
For $2 \in M,$ choose $(2,1,2,0)\in \mathcal{C}_2(\mathbb{D}).$ Add a node 5 to the network which demands $x_2$ and which has incoming edges from nodes 1 and 3.
\item
 For $1 \in M,$ choose $(1,2,2,0)\in \mathcal{C}_1(\mathbb{D}).$ Add a node 6 to the network which demands $x_1$ and which has incoming edges from nodes 2 and 3.
 \item
  For $2 \in M,$ choose $(2,1,0,2)\in \mathcal{C}_2(\mathbb{D}).$ Add a node 7 to the network which demands $x_2$ and which has incoming edges from nodes 1 and 4.
 \item
  For $1 \in M,$ choose $(1,2,0,2)\in \mathcal{C}_1(\mathbb{D}).$ Add a node 8 to the network which demands $x_1$ and which has incoming edges from nodes 2 and 4.
   \item
  For $1 \in M,$ choose $(1,0,2,2)\in \mathcal{C}_1(\mathbb{D}).$ Add a node 9 to the network which demands $x_1$ and which has incoming edges from nodes 3 and 4.
   \item
  For $2 \in M,$ choose $(0,1,2,2)\in \mathcal{C}_2(\mathbb{D}).$ Add a node 10 to the network which demands $x_2$ and which has incoming edges from nodes 3 and 4.
  \end{itemize}
  \begin{figure}[h]
\centering
\includegraphics[totalheight=3 in,width=2in]{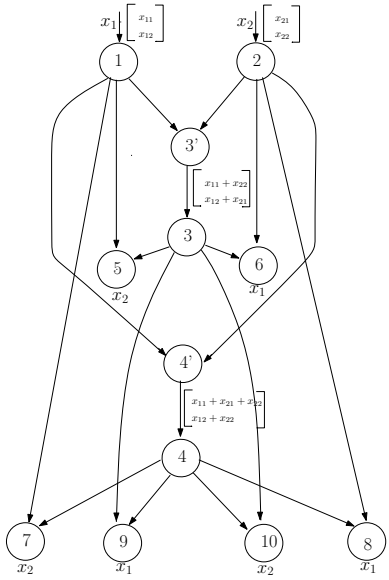}
\caption{The network constructed from the discrete polymatroid in Example 1}
\label{fig:network1}
\end{figure}

The network thus obtained is given in Fig. \ref{fig:network1}. A vector linear solution of dimension 2 over $\mathbb{F}_2$ shown in Fig. \ref{fig:network1} is obtained by taking the global encoding matrices for the edges $3' \rightarrow 3$ and $4' \rightarrow 4$ to be the matrices $A_3$ and $A_4$ given in Example 4. All the outgoing edges of a node which has in-degree one carry the same vector as that of the incoming edge. The network in Fig. \ref{fig:network1} does not admit a scalar linear solution over $\mathbb{F}_2$ as shown in the following lemma.
\begin{lemma}
The network given in Fig. \ref{fig:network1} does not admit a scalar linear solution over $\mathbb{F}_2.$
\begin{proof}
Observe that node 5 demands $x_2$ and the only path from 2 to 5 is via the edge $3' \rightarrow 3.$ Also, node 6 demands $x_1$ and the only path from 1 to 6 is via the edge $3' \rightarrow 3.$ To satisfy these demands, the edge $3' \rightarrow 3$ needs to carry $x_1+x_2.$ By a similar reasoning, to satisfy the demands of nodes 7 and 8, the edge $4' \rightarrow 4$ needs to carry $x_1+x_2.$ But if the edges $3' \rightarrow 3$ and $4' \rightarrow 4$ carry $x_1+x_2,$ the demands of nodes 9 and 10 cannot be satisfied.
\end{proof}
\end{lemma}

While the network in Fig. \ref{fig:network1} does not admit a scalar linear solution over $\mathbb{F}_2,$ it has a scalar linear solution over all fields of size greater than two, as shown in the following lemma.
\begin{lemma}
The network in Fig. \ref{fig:network1} admits a scalar linear solution over all fields of size greater than two.
\begin{proof}
It can be verified that the network shown in Fig. \ref{fig:network1} is matroidal with respect to the uniform matroid $U_{2,4}$ with the mapping $f$ from the edge set to the ground set $\lceil 4 \rfloor$ of the matroid defined as follows: for $i \in \lceil 4 \rfloor,$ all the elements of $In(i')$ are mapped to $head(i'),$  the elements of $out(i)$ and the edge joining $i'$ and $i$ are mapped to $i.$ Since $U_{2,4}$ is representable over all fields of size greater than or equal to three (follows from Proposition 6.5.2, Page 203, \cite{Ox}), the network in Fig. \ref{fig:network1} admits a scalar linear solution over all fields of size greater than two.
\end{proof}
\end{lemma} 

The network constructed in the previous example turned out to be matroidal with respect to a matroid representable over all fields other than $\mathbb{F}_2$ and as a result it admitted scalar linear solutions over all $\mathbb{F}_q$ other than $\mathbb{F}_2.$ In the following example, the constructed network is discrete polymatroidal with respect to a representable discrete polymatroid whereas it cannot be matroidal with respect to any representable matroid. Hence it is not scalar linearly solvable over any field, but is vector linear solvable.
\begin{example}
Let $V_i, i \in \lceil 12 \rfloor,$ denote the column span of the matrix $A_i$ shown in \eqref{matrix_M_network_1}. Let $\mathbb{D}$ denote the discrete polymatroid $\mathbb{D}(V_1,V_2,\dotso,V_{12}).$\\
\underline{\emph{Step 1:}} Choose $v=(2,2,2,2,0,0,0,0,0,0,0,0)$ from $\mathcal{R}(\mathbb{D})$ and it can be verified that $\vert (u)_{>0}\vert=4,\forall u \in \mathcal{R}(\mathbb{D}).$ Add nodes 1,2,3 and 4 to the network which generates messages $x_1,x_2,x_3,x_4.$ Set $T=M=\{1,2,3,4\}.$\\
\underline{\emph{Step 2:}}
\begin{itemize}
\item
Choose $(2,2,0,0,1,0,0,0,0,0,0,0)\in\mathcal{C}_5(\mathbb{D}).$ Add node $5'$ to the network with incoming edges from 1 and 2. Also, add node 5 with an edge from $5'.$ $T=\{1,2,3,4,5\}.$
\item
Choose $(2,2,0,0,0,0,1,0,0,0,0,0)\in\mathcal{C}_7(\mathbb{D}).$ Add node $7'$ to the network with incoming edges from 1 and 2. Also, add node 7 with an edge from $7'.$ $T=\{1,2,3,4,5,7\}.$
\item
Choose $(2,2,0,0,0,0,0,1,0,0,0,0)\in\mathcal{C}_8(\mathbb{D}).$ Add node $8'$ to the network with incoming edges from 1 and 2. Also, add node 8 with an edge from $8'.$ $T=\{1,2,3,4,5,7,8\}.$
\item
Choose $(2,0,0,0,0,0,2,0,1,0,0,0)\in\mathcal{C}_9(\mathbb{D}).$ Add node $9'$ to the network with incoming edges from 1 and 7. Also, add node 9 with an edge from $9'.$ $T=\{1,2,3,4,5,7,8,9\}.$
\item
Choose $(0,0,0,0,0,0,2,0,2,1,0,0)\in\mathcal{C}_{10}(\mathbb{D}).$ Add node $10'$ to the network with incoming edges from 7 and 9. Also, add node 10 with an edge from $10'.$ $T=\{1,2,3,4,5,7,8,9,10\}.$
\item
Choose $(0,2,0,0,0,1,0,0,2,0,0,0)\in\mathcal{C}_6(\mathbb{D}).$ Add node $6'$ to the network with incoming edges from 2 and 9. Also, add node 6 with an edge from $6'.$ $T=\{1,2,3,4,5,6,7,8,9,10\}.$
\item
Choose $(0,0,0,0,0,2,0,0,2,0,1,0)\in\mathcal{C}_{11}(\mathbb{D}).$ Add node $11'$ to the network with incoming edges from 6 and 9. Also, add node 11 with an edge from $11'.$ $T=\{1,2,3,4,5,7,8,9,10,11\}.$
\item
Choose $(0,0,0,0,0,0,0,0,0,2,2,1)\in\mathcal{C}_{12}(\mathbb{D}).$ Add node $12'$ to the network with incoming edges from 10 and 11. Also, add node 12 with an edge from $12'.$ $T=\{1,2,3,4,5,7,8,9,10,11,12\}.$
\end{itemize}
\underline{\emph{Step 3:}}
\begin{itemize}
\item
For $1 \in M,$ choose $(1,0,0,0,2,2,0,0,0,0,0,0) \in \mathcal{C}_1(\mathbb{D}).$ Add a node 13 which demands $x_1$ and has incoming edges from nodes 5 and 6.
\item
For $1 \in M,$ choose $(1,0,0,0,2,0,0,0,0,2,0,0) \in \mathcal{C}_1(\mathbb{D}).$ Add a node 14 which demands $x_1$ and has incoming edges from nodes 5 and 10.
\item
For $1 \in M,$ choose $(1,0,0,0,2,0,0,0,2,0,0,0) \in \mathcal{C}_1(\mathbb{D}).$ Add a node 15 which demands $x_1$ and has incoming edges from nodes 5 and 9.
\item
For $1 \in M,$ choose $(0,1,0,0,2,2,0,0,0,0,0,0) \in \mathcal{C}_2(\mathbb{D}).$ Add a node 13 which demands $x_2$ and has incoming edges from nodes 5 and 6.
\item
For $1 \in M,$ choose $(0,1,0,0,2,0,0,0,0,0,2,0) \in \mathcal{C}_2(\mathbb{D}).$ Add a node 14 which demands $x_2$ and has incoming edges from nodes 5 and 11.
\item
For $1 \in M,$ choose $(0,1,0,0,2,0,0,0,0,0,0,2) \in \mathcal{C}_2(\mathbb{D}).$ Add a node 15 which demands $x_2$ and has incoming edges from nodes 5 and 12.
\item
For $4 \in M,$ choose $(0,0,0,1,0,0,0,2,0,0,0,2) \in \mathcal{C}_4(\mathbb{D}).$ Add a node 16 which demands $x_4$ and has incoming edges from nodes 8 and 12.
\item
For $3 \in M,$ choose $(0,0,1,0,0,0,0,2,0,0,2,0) \in \mathcal{C}_3(\mathbb{D}).$ Add a node 14 which demands $x_3$ and has incoming edges from nodes 8 and 11.
\item
For $3 \in M,$ choose $(0,0,1,0,0,0,2,2,0,0,0,0) \in \mathcal{C}_3(\mathbb{D}).$ Add a node 15 which demands $x_3$ and has incoming edges from nodes 7 and 8.
\item
For $4 \in M,$ choose $(0,0,0,1,0,0,0,2,0,2,0,0) \in \mathcal{C}_4(\mathbb{D}).$ Add a node 13 which demands $x_4$ and has incoming edges from nodes 8 and 10.
\item
For $3 \in M,$ choose $(0,0,1,0,0,0,0,2,2,0,0,0) \in \mathcal{C}_3(\mathbb{D}).$ Add a node 14 which demands $x_3$ and has incoming edges from nodes 8 and 9.
\item
For $4 \in M,$ choose $(0,0,0,1,0,0,2,2,0,0,0,0) \in \mathcal{C}_4(\mathbb{D}).$ Add a node 15 which demands $x_4$ and has incoming edges from nodes 7 and 8.
\end{itemize}

The network thus constructed is shown in Fig. \ref{fig:network2}. The vector linear solution of dimension 2, which is in fact a vector routing solution, is obtained by choosing the global encoding matrix of the edge $i' \rightarrow i,i \in \lceil 12 \rfloor,$ to be $A_i$, as shown in Fig. \ref{fig:network2}. All the outgoing edges of a node which has in-degree one carry the same vector as that of the incoming edge.
 \begin{figure}[t]
\centering
\includegraphics[totalheight=3 in,width=3in]{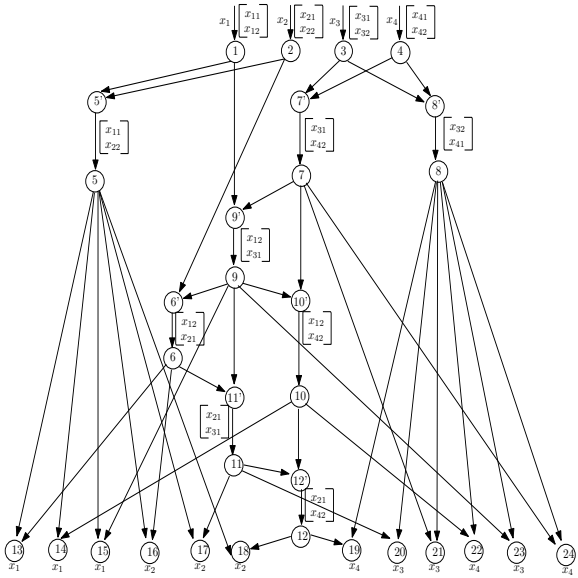}
\caption{A network which is vector linearly solvable but not scalar linearly solvable}
\label{fig:network2}
\end{figure}
\end{example}
The following lemma shows that the network in Fig. \ref{fig:network2} is not scalar linearly solvable.
\begin{lemma}
The network in Fig. \ref{fig:network2} is not scalar linearly solvable.
\begin{proof}
To prove the lemma, it is shown that the network cannot be matroidal with respect to a representable matroid. On the contrary, assume that the network is matroidal with respect to a representable matroid $\mathbb{M}=(\lceil n \rfloor,\rho)$ and let $f$ be the network-matroid mapping. Let the set of one dimensional vector spaces $V_i, i \in \lceil n \rfloor$ form a representation of $\mathbb{M}.$ All the outgoing edges of a node which has in-degree one carry the same vector as that of the incoming edge. For simplicity, let $i$ denote the incoming edge of node $i,$ where $i \in \lceil 12 \rfloor.$ Let $g(x)=\rho(f(x)), x \subseteq \lceil 12 \rfloor.$ 

We have $g(\{1,2\})\leq 2.$ From (DN2), it follows that $\sum_{i \in \lceil 4 \rfloor}\epsilon_{f(i)} \in \mathcal{I}(\mathbb{M}).$ Hence we have $\sum_{i \in \lceil 2 \rfloor}\epsilon_{f(i)} \in \mathcal{I}(\mathbb{M}),$ from which it follows that $2 \leq g(\{1,2\}).$ Hence, we have $g(\{1,2\})=2.$ Similarly, we also have $g(\{3,4\})=2.$

It is claimed that $g(\{5\})=1.$ Otherwise, $g(\{5\})$ has to be 0. In that case, since the nodes 13 and 16 demand $x_1$ and $x_2$ respectively, from (DN3) it follows that $dim(V_{f(1)}+V_{f(6)})=dim(V_{f(6)})$ and $dim(V_{f(2)}+V_{f(6)})=dim(V_{f(6)}).$ Hence $V_{f(1)}=V_{f(2)}$ which is not possible and hence $g(\{5\})$ has to be 1. Similarly, it can be shown that $g(\{8\})=1.$
We have,
\begin{align}
\label{eqn_first}
g(\lbrace 3,8 \rbrace)+g(\{4,8\}) &\geq g(\lbrace 8 \}))+g(\{3,4,8\})\\
\label{eqn1}
 &\geq 1+ g(\{3,4\})=3,
\end{align}
where \eqref{eqn_first} holds since $g(\{3,4,8\})=g(\{4,8 \})$ (follows from (DN3)) and \eqref{eqn1} follows from the facts that $g(\{8\})=1$ and $g(\{3,4\})=2.$

Similarly, it can be shown that  
\begin{align}
\label{eqn4}
g(\lbrace 1,5 \rbrace)+g(\{2,5\})\geq 3.
\end{align}
Also, we have,
\begin{align}
\label{eqn_third}
g(\{2,5\})+g(\{3,8\})&=g(\{2,5,3,8\})\\
\nonumber
&\leq g(\{2,5,3,8,11 \})\\
\label{eqn_fourth}
&\hspace{-1 cm}\leq g(\{2,5,11\})+g(\{3,8,11\})-g(\{11\})\\
\label{eqn2}
&\hspace{-1 cm}=g(\{5,11\})+g(\{8,11\})-1\leq 3,
\end{align}
where \eqref{eqn_third} follows from the fact that $dim\left(V_{f(1)}+V_{f(5)}\right)+dim\left(V_{f(3)}+V_{f(8)}\right)=dim\left(V_{f(1)}+V_{f(5)}+V_{f(3)}+V_{f(8)}\right)$ and \eqref{eqn_fourth} follows from (D2).
Similarly, it can be shown that 
\begin{align}
\label{eqn3}
g(\{2,5\})+g(\{4,8\})\leq 3.
\end{align}
From \eqref{eqn1}, \eqref{eqn2} and \eqref{eqn3}, we get $g(\{2,5\})\leq 1.5.$ Similarly, it can be shown that $g(\{1,5\}) \leq 1.5.$ Hence, from \eqref{eqn4}, we get $g(\{1,5\})=g(\{2,5\})=1.5$ which is not an integer, resulting in a contradiction. Hence, the network in Fig. \ref{fig:network2} cannot be matroidal with respect to any representable matroid.
\end{proof}
\end{lemma}
\section{discussion}
The connection between the vector linear solvability of networks over a field $\mathbb{F}_q$ and the representation of discrete polymatroids was established. It was shown that for a network, a vector linear solution over a field $\mathbb{F}_q$ exists if and only if the network is discrete polymatroidal with respect to a representable discrete polymatroid. An algorithm to construct networks from discrete polymatroids was provided. Sample constructions of networks from representable discrete polymatroids which have vector linear solutions but no scalar linear solution over $\mathbb{F}_q$ were provided.   
   
\end{document}